\documentclass[journal,comsoc]{IEEEtran}
\usepackage[T1]{fontenc}
\usepackage{graphicx}

\graphicspath{ {images/} }
\usepackage{algorithmic}
\usepackage{algorithm}
\usepackage{amsmath}
\usepackage{amssymb}
\usepackage{textcomp}
\usepackage{amsthm}
\usepackage{wrapfig}
\usepackage{lipsum}
\usepackage{xcolor}
\graphicspath{ {images/} }
\usepackage{array}
\ifCLASSOPTIONcompsoc
  \usepackage[caption=false,font=normalsize,labelfont=sf,textfont=sf]{subfig}
\else
  \usepackage[caption=false,font=footnotesize]{subfig}
\fi
\usepackage{stfloats}
\usepackage{url}
\usepackage{balance}

\usepackage[cmintegrals]{newtxmath}

\begin{document}
\title{Non-Stationary Wireless Channel Modeling Approach Based on Extreme Value Theory for Ultra-Reliable Communications}

\author{Niloofar~Mehrnia,~\IEEEmembership{Student Member,~IEEE,}
        Sinem~Coleri,~\IEEEmembership{Senior Member,~IEEE}
\thanks{Niloofar Mehrnia and Sinem Coleri are with the Department of Electrical and Electronics Engineering,
Koc University, Istanbul, Turkey (e-mail: nmehrnia17@ku.edu.tr; scoleri@ku.edu.tr).}
\thanks{Niloofar Mehrnia is also with Koc University Ford Otosan Automotive Technologies Laboratory (KUFOTAL), Sariyer, Istanbul, Turkey, 34450.}
\thanks{Sinem Coleri acknowledges the support of Ford Otosan.}
}

\maketitle

\begin{abstract}
A proper channel modeling methodology that characterizes the statistics of extreme events is key in the design of a system at an ultra-reliable regime of operation. The strict constraint of ultra-reliability corresponds to the packet error rate (PER) in the range of $10^{-9}-10^{-5}$ within the acceptable latency on the order of milliseconds. Extreme value theory (EVT) is a robust framework for modeling the statistical behavior of extreme events in the channel data. In this paper, we propose a methodology based on EVT to model the extreme events of a non-stationary wireless channel for the ultra-reliable regime of operation. 
This methodology includes techniques for splitting the channel data sequence into multiple groups concerning the environmental factors causing non-stationarity, and fitting the lower tail distribution of the received power in each group to the generalized Pareto distribution (GPD). The proposed approach also consists of optimally determining the time-varying threshold over which the tail statistics are derived as a function of time, and assessing the validity of the derived Pareto model. Finally, the proposed approach chooses the best model with minimum complexity that represents the time variation behavior of the non-stationary channel data sequence. Based on the data collected within the engine compartment of Fiat Linea under various engine vibrations and driving scenarios, we demonstrate the capability of the proposed methodology in providing the best fit to the extremes of the non-stationary data. The proposed approach significantly outperforms the channel modeling approach using the stationary channel assumption in characterizing the extreme events.     

\end{abstract}


\IEEEpeerreviewmaketitle

\section{Introduction}
\label{sec:intro}
Ultra-reliable and low latency communication (URLLC) is one of the key features of beyond fifth-generation (5G) networks, with the potential to support a vast set of mission-critical applications in vehicular communications, remote surgeries, and virtual reality \cite{interface_01}-\cite{urllc_02}. Under the constraints of a URLLC service, the reliability is defined as achieving the packet error rate (PER) in the range of $10^{-9}-10^{-5}$ with the latency on the order of milliseconds. An accurate channel modeling quantifying the tail statistics produced by extreme events is needed to satisfy the requirements of the URLLC systems.

Existing studies on channel modeling for URLLC can be categorized into two folds: One provides an extrapolation-based framework that extends the applicability of the available practical channel models in the ultra-reliability relevant regime \cite{urllc_02}; The other proposes new performance measures for the ultra-reliability of the channel \cite{urllc_07}\nocite{urllc_08}-\cite{urllc_05}. However, none of these frameworks propose a channel modeling methodology to derive and verify ultra-reliability statistics. In \cite{Mehrnia2020}, we propose a novel channel modeling methodology for URLLC based on extreme value theory (EVT) to statistically derive the lower tail of the received power by characterizing the probabilistic distribution of extreme events happening rarely. However, the channel was assumed stationary, whereas, in reality, the wireless channel statistics vary over time due to the dynamic environment. Hereupon, the usage of techniques for the estimation of the time-varying parameters of the non-stationary channel data sequence is required.

EVT is a powerful framework characterizing the probabilistic distribution of infrequent extreme events or equivalently the tail distribution. EVT has been recently utilized at the data link and network layers to model the tail statistics of queue length and delay \cite{upperlayer_03}-\cite{upperlayer_05}. Additionally, EVT has been employed to derive closed-form asymptotic expressions for the throughput, bit error rate (BER), and PER over different fading channels \cite{upperlayer_04}\nocite{upperlayer_02}-\cite{upperlayer_01}. However, these data link and network layer studies use the extrapolation of existing average statistics-based channel models in the ultra-reliable region, which have not yet been proven to be accurate experimentally. 
EVT has also been utilized at physical layer for wireless channel modeling to provide a fit to the whole distribution of large/small-scale fading \cite{fadingevt_01}\nocite{fitevt_01}\nocite{fitevt_02}\nocite{fitevt_03}-\cite{fitevt_04}. Nevertheless, EVT has never been incorporated into wireless channel modeling to estimate the tail statistics and address the constraint in the ultra-reliability region. This constraint requires the development of methodologies for the determination of the optimum threshold below/above which the samples correspond to the extreme events, and acquiring a large number of samples to capture the extreme events occurring rarely. \cite{Mehrnia2020} is the only existing study focusing on modeling the statistics of the channel model for URLLC by using the concept of EVT, however, stationary channel assumption may not be realistic for channels with time-varying characteristics. 

The goal of this paper is to propose a novel channel modeling methodology based on EVT for the statistical characterization of the lower tail distribution of a non-stationary channel data. We use received signal power at constant transmit power as channel data, since the received signal power is equivalent to using the squared amplitude of the channel state information \cite{urllc_02}, \cite{Mehrnia2020}. This paper extends the EVT based wireless channel modeling methodology in \cite{Mehrnia2020} for non-stationary channel data by determining the measurable external factors that cause non-stationarity. Then, it models the parameters of fitting distributions as a change-point function of time. First, the wireless channel data sequence is split into smaller groups, each of which corresponds to the same value of the external factor causing non-stationarity. Then, the generalized Pareto distribution is fitted to the lower tail of the received power distribution in each group. The original contributions of this work are listed as follows:
\begin{itemize}
    \item We provide a novel methodology for determining the measurable external factors that cause non-stationarity in the channel data sequence. This methodology splits the data sequence into smaller stationary groups, each of which contains the samples collected under the same factor.
    \item We provide a novel approach for modeling the extremes of each group exceeding a given threshold by fitting the generalized Pareto distribution (GPD) to the distribution of the channel lower tail while the threshold is determined optimally. Thereupon, we model the shape and scale parameters of GPD as a change-point function of time. 
    \item We demonstrate the superiority of the proposed channel modeling methodology for deriving the tail statistics of a non-stationary channel sequence in terms of the deviance statistic, compared to the case in which the channel is assumed stationary. 
\end{itemize}

The rest of the paper is organized as follows: Section~\ref{sec:background} describes the basics of EVT for stationary and non-stationary sequences. Section~\ref{sec:methodology} presents the channel modeling methodology for characterizing the extremes of a non-stationary sequence. Section \ref{sec:performance evaluation} provides the performance evaluation of the proposed algorithm on the data collected within the engine compartment of Fiat Linea under various engine vibrations and driving scenarios. Finally, Section~\ref{sec:conclusions} concludes the paper.

\section{Background}\label{sec:background}

\subsection{Extreme Value Theory for Stationary Sequences}
EVT provides a robust framework for analyzing the statistics of extreme events happening rarely through modeling the probabilistic distribution of the values exceeding a given threshold by using the GPD. Assume that $\{x_1,...,x_N\}$ is an independent and identically distributed stationary sequence, where $x_{i}$ denotes the $i^{th}$ received power for $i \in \{1,...,N\}$. Then, according to the EVT, the tail distribution of the power sequence, i.e., the probabilistic distribution of the power values exceeding a given threshold $u$, can be expressed as
\begin{equation*}
\label{eqn:gpd dist}
    G_{u}(y) = 1-\Bigg[1+\frac{\xi y}{\tilde{\sigma}_{u}}\Bigg]^{-1/\xi},
\end{equation*}
where $y$ is a non-negative value denoting the exceedance below threshold $u$, i.e., ($y=u-X$), $X$ denotes any $x_i$ below threshold $u$; $G_{u}(y)$ is in the form of the GPD; and \textcolor{black}{$\xi$ and $\tilde{\sigma}_{u}=\sigma+\xi(u-\mu)$ are shape and scale parameters of the GPD, respectively. Here, $\mu$ and $\sigma$ are the location and scale parameters of the generalized extreme value (GEV) distribution fitted to the CDF of $m_N = min \{ x_1,...,x_N\}$, respectively \cite[Theorem~1]{Mehrnia2020}}, \cite{evt_11}. 

\subsection{Extreme Value Theory for Non-Stationary Sequences}
In a non-stationary sequence, the threshold, shape, and scale parameters of the GPD model are time-varying. Therefore, the distribution of the values exceeding a given threshold is modeled by using a general Pareto model with time-varying parameters given by
\begin{equation}
    \left(u_{\left(t\right)}-X_t\middle| X_t<u_{\left(t\right)}\right) \sim GPD\left(\tilde{\sigma}_{\left(u,t\right)},\xi_{\left(t\right)}\right),
\end{equation}
where $u_{(t)}$, $\tilde{\sigma}_{(u,t)}$, and $\xi_{(t)}$ are the time-varying threshold, scale and shape parameters of the generalized Pareto distribution, respectively.

Change-point approach is utilized to model the impact of time on the extremes of a non-stationary sequence through a time-varying factor. In change-point technique, the non-stationary sequence is broken into smaller stationary groups, each of which corresponds to the same external factor. In this technique, the parameters of the GPD model fitted to the extremes of the stationary groups are assumed to be fixed as long as the external factor is constant. The time varying parameters are then expressed as
\begin{align}
\label{eqn:changepoint}
\theta_{(t)} = \left\{
    \begin{array}{ll}
     \theta_1,   &  0<t \leq t_1 \\
     \theta_2,   &  t_1<t \leq t_2\\
     .      & . \\
     \theta_M,   &  t_{M-1}<t \leq t_M
    \end{array} \right.
\end{align}
for $0<t \leq t_M$, where $\theta_{(t)}$ is the time varying $u_{(t)}$, $\tilde{\sigma}_{(u,t)}$, or $\xi_{(t)}$; $\theta_m$ is the determined constant $u$, $\tilde{\sigma}_{(u)}$, or $\xi$ for group $m \in \{1,..., M\}$; and $M$ is the number of groups between times $0$ and $t_M$.

\section{Methodology}
\label{sec:methodology}
The objective of the proposed methodology is to model a non-stationary channel at an ultra-reliable regime of operation based on EVT. The main features of the suggested algorithm are as follows:
\begin{enumerate}
\color{black}
    \item We perform the Augmented Dickey-Fuller (ADF) test on the data to check if it is non-stationary. The test is based on the fact that the mean and variance of a stationary time series do not change over time \cite{ADFtest}. 
    \color{black}
    \item If the channel data sequence is not stationary according to the ADF test result, we determine the external factors under which the parameters of the GPD are changing and categorize the sequence into $M$ different groups.
    \item We model the tail distribution of the received power in each group by using GPD. 
    \item We select the simplest GPD model with enough accuracy by using the deviance statistic $D$.
\end{enumerate}

\begin{figure*}[ht] 
\centering{
\includegraphics[scale=0.62]{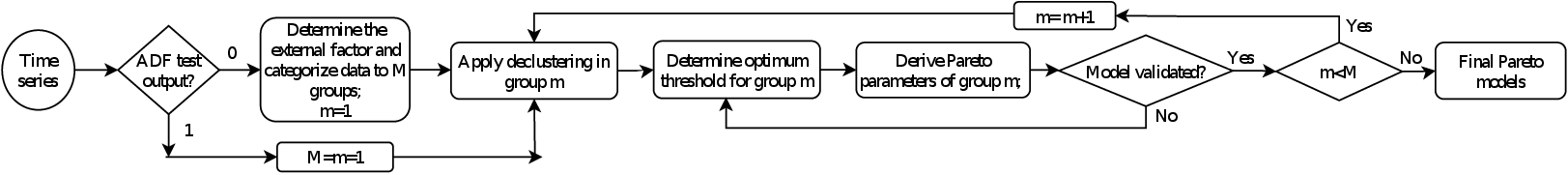}
\caption{Flowchart of the proposed channel modeling framework for a non-stationary process.}
\label{fig:maindiagram}}
\end{figure*}

The proposed approach starts by applying ADF test on the channel data to check the stationarity. If the ADF test logical result is $0$, we do not have enough evidence to reject the null hypothesis of non-stationary assumption. The logical result is $0$ if $p$-value is less than $\alpha$, where $p$-value is the index measuring the strength of the evidence against the null hypothesis $H_0$, and $\alpha$ is defined as the probability of rejecting the null hypothesis when the null hypothesis is true, i.e., probability of making a wrong decision. Then, the samples are categorized into $M$ groups based on the factors that cause non-stationarity. If the ADF test logical result is $1$, we reject the null hypothesis and consider the channel as stationary. Afterwards, since the input of EVT is necessarily a sequence of i.i.d. samples, we remove the time-dependency between the samples of each group by using declustering approach. Then, EVT is applied to obtain the optimum threshold under which the received powers are considered extremely low values. Hereupon, we fit GPD with the associated shape and scale parameters to the values below the threshold to model the tail of the receive power channel data.  Finally, we obtain the log-likelihoods of the GPD models under different factors to select the model with the lowest complexity, i.e., minimum total number of scale and shape parameters, based on the deviance statistic. The proposed algorithm is depicted in Fig.~\ref{fig:maindiagram} and explained in detail next.

\subsection{Modeling of Tail Distribution by using GPD}\label{sec:model tail}
To model the tail distribution of the received power in each group, first, the sequence of measured samples is converted into an i.i.d. sequence by removing their dependency. The threshold exceedances in the observation sequence are inherently time dependent since one extremely low power is likely to be followed by another. On the other hand, since the input of EVT needs to be a sequence of i.i.d random variables, we use declustering approach to remove time dependencies \cite{Mehrnia2020}, \cite{evt_04}. Then, EVT is applied to the sequence of i.i.d. samples for determining the optimum threshold and estimating the parameters of the Pareto distribution by using the maximum likelihood estimator (MLE). The optimum threshold is determined by applying two complementary methods, mean residual life (MRL) and parameter stability methods, in EVT. Next, the validity of the Pareto model corresponding to the optimum threshold is assessed by using probability plots. 

The time-dependency of the observed samples is removed via the declustering approach, where the samples in each individual group are divided into multiple clusters. Each cluster includes consecutive dependent observations and the clusters are separated by a certain sample gap to ensure the independency between clusters \cite{Mehrnia2020}. In this method, first, we assume a threshold $u$ and look for the first sample $x_i$ below this threshold to initiate the first cluster. Upon observing the first $x_i<u$, $x_i$ and all its consecutive samples below $u$ will be assigned to the first cluster. Once a sample over $u$ is detected, we let the cluster to continue for $r$ more successive values and then, terminate the cluster, if no value below $u$ is observed. The next cluster starts with the next value below the threshold $u$. In the second step of the declustering method, we extract the minimum value of each cluster, apply EVT to the cluster minima, and model their tail distribution by using GPD.      

Optimum threshold determination is performed by using MRL and parameter stability methods. MRL method states that if we determine the mean value of samples exceeding a given threshold $u$ and then, plot mean excesses, i.e., $E(u-X|X<u)$, against the threshold, the optimum threshold is the highest threshold below which, mean excess is a linear function of $u$. Though the MRL method is applied to the data sequence prior to the estimation of the Pareto model parameters with less complexity than the parameter stability method, it is sometimes difficult to obtain the optimum threshold explicitly. Therefore, it is required to use the complementary parameter stability method. Parameter stability method states that if we fit GPD to the values exceeding a given threshold for a variety of the thresholds and extract the corresponding Pareto parameters and then plot parameters against the threshold $u$, the optimum threshold is the highest threshold below which the estimated shape and modified scale parameters are linear with respect to $u$. Here,  modified scale parameter is defined as $\sigma^{*} = \tilde\sigma_{u} - \xi u$ and the linearity relation is assessed by using the R-squared statistical measure, denoted by $R^{2}$ \cite{Mehrnia2020}.

The validity of the GPD model is assessed by using probability plots. Probability plots, consisting of Probability/Probability (PP) plot and Quantile/Quantile (QQ) plot, are graphical techniques used to assess the validity of the models fitted to the empirical values. In the PP plot, we compare the empirical probability of occurrence for an extreme value with the corresponding probability obtained by the GPD, while in the QQ plot, we compare the empirical extreme quantile with the corresponding value obtained by the inverse of GPD. If the GPD appropriately models the extreme values exceeding threshold $u$, then, both PP and QQ plots fit the unit diagonal line, i.e., the $45^\circ$ line \cite{evt_04}, \cite{evt_01}. PP plot consists of the pairs 
\begin{align}
\label{eqn:probilityplot}
\begin{split}
   \{ (\frac{i}{k+1}), G_{u}(y_{i})\};
   G_{u}(y_{i}) = 1 - \Bigg( 1-\frac{{\xi}y_{i}}{\tilde{\sigma}_{u}} \Bigg) ^{-1/{\xi}},
\end{split}
\end{align}
where $y_{i}$, $i\in\{1,...,k\}$, is the absolute value of the difference between the threshold $u$ and $x_{i}$ sample exceeding the threshold; $k$ is the number of values exceeding threshold $u$; $G_{u}(.)$ is the Pareto model fitted to the tail distribution; and $\xi$ and $\tilde{\sigma}_u$ are the associated shape and scale parameters, respectively \cite{evt_04}. On the other hand, QQ plot consists of the pairs 
\begin{align}
    \label{eqn:quantileplot}
    \begin{split}
        \{x_{i},({G}_{u}^{-1}(\frac{i}{k+1})\};
        {G}_{u}^{-1}(z_{i}) = u - \frac{\tilde{\sigma}_{u}}{\xi}\bigg[1 - z_{i}^{-{\xi}} \bigg],
    \end{split}
\end{align}
where $x_{i}$, $i\in\{1,...,k\}$, is the $i^{th}$ extreme quantile exceeding threshold $u$; ${G}_{u}^{-1}(.)$ is the inverse of the Pareto model fitted to the tail distribution; $z_{i}$ is the probability of occurrence associated with the extreme quantile $x_{i}$; and $k$, $\xi$ and $\tilde{\sigma}_u$ are same as those defined in Eqn.~(\ref{eqn:probilityplot}). 

\subsection{Modeling Time-Varying Parameters of GPD}\label{sec:time-varying parameters}
To model the extreme values in a non-stationary sequence, upon determining the external factors that affect Pareto parameters, we determine an optimum threshold and the corresponding Pareto model parameters for each group.
The time-varying parameters are modeled according to the change-point expression given in Eqn.~(\ref{eqn:changepoint}).

\subsection{Final Model Selection}\label{sec:deviance}
With the possibility of modeling any combination of the GPD parameters as functions of time, there is a catalog of models to choose from. The basic principle is parsimony, obtaining the simplest model that explains as much of the variation in the data as possible. In this regard, we obtain the log-likelihood of the GPD associated with each external factor in Section~\ref{sec:time-varying parameters} as 
\begin{multline}
\label{eqn:likelihood}
    l_{GPD} = -\Big[ \sum_{t=1}^{N} \log{\tilde{\sigma}_{(u,t)}} +
    \big(1+\frac{1}{\xi_{(t)}}\big) \log {(1+ \frac{\xi_{(t)} y_{(t)}}{\tilde{\sigma}_{(u,t)}}})
    \Big],
\end{multline}
where $N$ is the maximum number of the samples; $\tilde{\sigma}_{(u,t)}$ and $\xi_{(t)}$ are the time varying scale and shape parameters at time $t$, respectively; and $y_{(t)}=u_{(t)}-x_{(t)}$ is the absolute value of the difference between the threshold $u_{(t)}$ and the sample $x_{(t)}$ exceeding $u_{(t)}$ at time $t$. Finally, we apply the deviance statistic test $D=2\{l_{GPD_1}-l_{GPD_0}\}$, where $l_{GPD_1}$ and $l_{GPD_0}$ are the log-likelihoods of $GPD_1$ and $GPD_0$ models, respectively. The primarily model $GPD_0$ is rejected by the test at the $\alpha$-level of significance if $D>c_\alpha$, where $c_\alpha$ is the $\left(1-\alpha\right)$ quantile of the $\chi_k^2$ distribution for a small $\alpha$ value between $0.01$ and $0.1$; and $k$ in $\chi_k^2$ is the dimensionality difference, i.e., difference between the total number of scale and shape parameters in each model. 

\section{Performance Evaluation}
\label{sec:performance evaluation}
The goal of this section is to evaluate the performance of the proposed methodology in modeling the non-stationary channel for URLLC, and choosing the best GPD fitted to the non-stationary sequence according to the deviance statistic test.

\begin{figure}[ht] 
\centering{
\includegraphics[width=0.63\columnwidth, height=3cm]{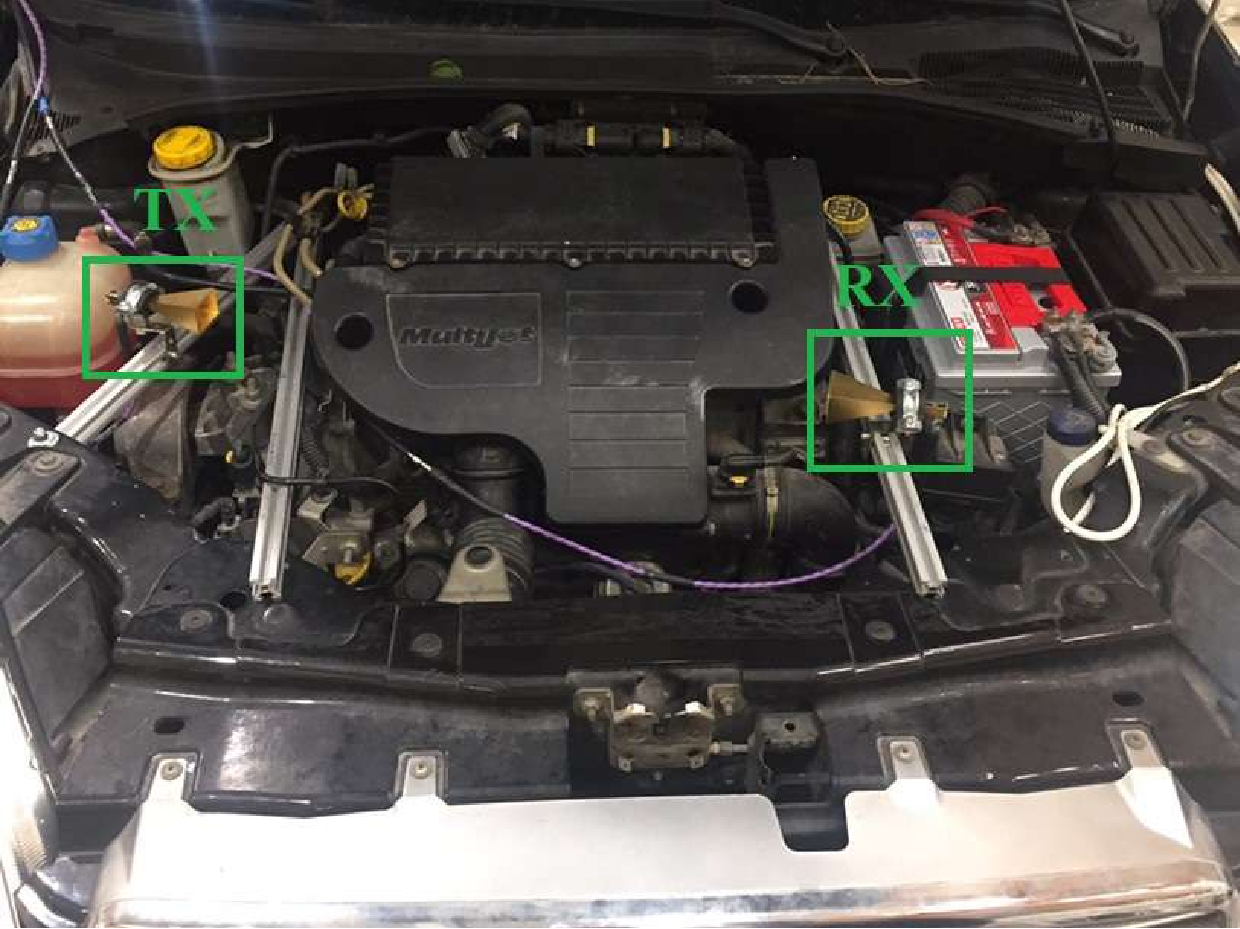}
\caption{Measurement setup with the transmitter (TX) and receiver (RX) antennas located in the engine compartment of Fiat Linea.}
\label{fig:setup}}
\end{figure}

The measured channel data were collected within the engine compartment of Fiat Linea at $60$ GHz under different driving scenarios and road conditions, including the static car, driving on a smooth road, and ramp road. The antennas within the engine compartment are located such that the effect of the engine vibration is observed in the received power, as shown in Fig.~\ref{fig:setup}. A Vector Network Analyzer (VNA) (R$\And$S$\textsuperscript{\textregistered}$ ZVA$67$) is connected to the transmitter and receiver via the R$\And$S$\textsuperscript{\textregistered}$ ZV-Z$196$ port cables with maximum $4.8$ dB transmission loss. The horn transmitter and receiver antennas with a nominal $24$ dBi gain and $12^\circ$ vertical beam-width operate at $50$-$75$ GHz. We have captured about $10^{6}$ successive samples for $30$ minutes with a time resolution of $2$ ms. We use \textsc{MATLAB} for the implementation of the proposed algorithm.

\subsection{External Factors Causing Non-Stationarity}\label{sec:param_nonstat}
We have identified the dependence of the parameters that cause non-stationarity on both the driving scenario and the quality of the road assessed by the driver of the vehicle. Therefore, for the collected channel data, we define the parameter taking three discrete values corresponding to the following three groups: static car, smooth road, and ramp road.

\subsection{Modeling Time-Varying Parameters of GPD Within the Engine Compartment}\label{sec:engin_timevarying_param}
\begin{figure}[h]
\centering
\captionsetup[subfigure]{labelformat=empty}
     \begin{center}
        \subfloat[(a)]{
            \label{fig:meanresvsth}
            \includegraphics[width=0.47\columnwidth]{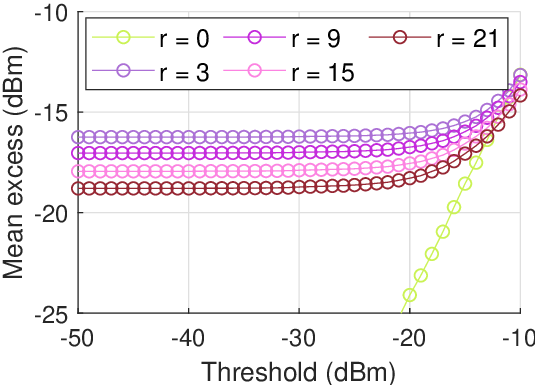}
        }%
        \subfloat[(b)]{
            \label{fig:modscalevsth}
            \includegraphics[width=0.47\columnwidth]{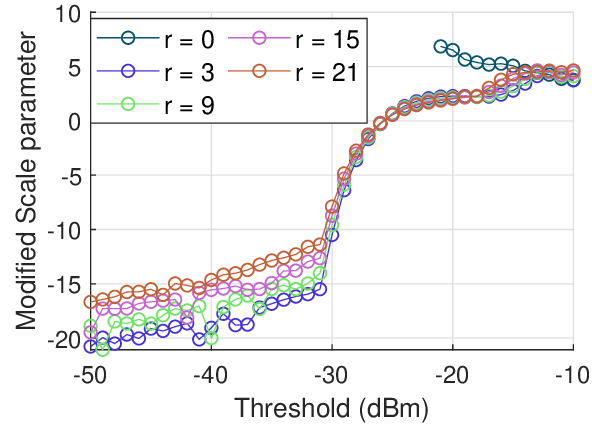}
        }\\
         \subfloat[(c)]{
            \label{fig:shapevsth}
            \includegraphics[width=0.47\columnwidth]{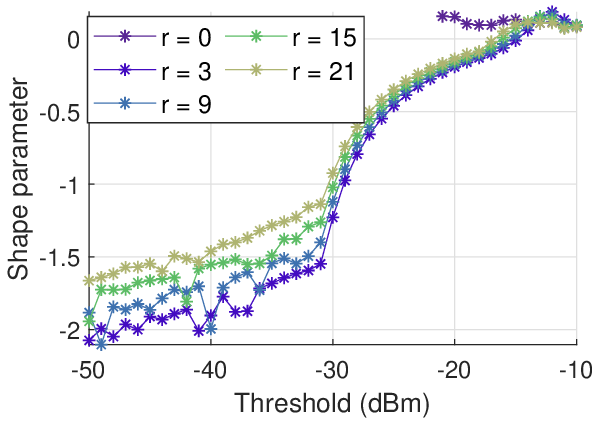}
            }
            \subfloat[(d)]{
            \label{fig:ppqq}
            \includegraphics[width=0.46\columnwidth]{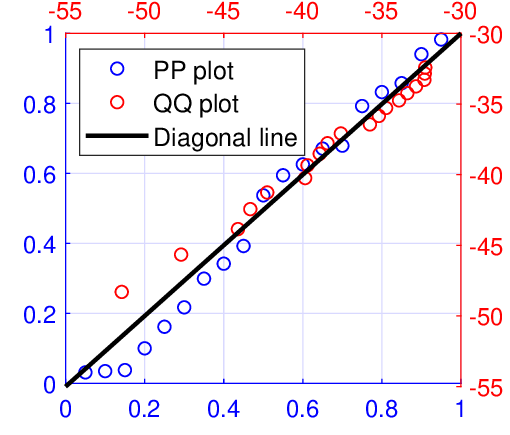}
 }\\
    \end{center}
    \caption{Mean excesses, shape, and modified scale parameters of GPD at different thresholds and minimum gaps between the clusters for the ramp road group: (a) Mean excess, (b) Modified scale parameter, (c) Shape parameter, and (d) Probability plots where the black line is a diagonal line, i.e., $y=x$, assessing the goodness of fit; $x$/$y$ axes are empirical/modeled values.}%
   \label{fig:paramsvsth}
\end{figure}

Fig.~\ref{fig:paramsvsth} shows the mean excess, shape and modified scale parameters of the GPD fitted to the filtered i.i.d. received power samples at different thresholds and minimum gaps between the clusters, as well as the probability plots for the group of ramp road. Fig.~\ref{fig:meanresvsth} illustrates the MRL plot, i.e., the mean of samples exceeding a given threshold, where the threshold varies between $-50$ dBm and $-10$ dBm. When the minimum gap between the clusters $r$ is $0$, the MRL plot linearly increases as threshold increases at all threshold values, thus, the optimum threshold cannot be recognized. However, by choosing $r>0$, the MRL changes linearly in $u$ only for $u<-20$ dBm, denoting that $-20$ dBm is the largest threshold value for which $R^{2}$ is greater than $0.95$ for all $r>0$.
On the other hand, Figs.~ \ref{fig:modscalevsth} and \ref{fig:shapevsth} show the shape and modified scale parameters of the fitted GPD model for the parameter stability method. These figures illustrate that by choosing $r>15$, the estimated parameters of GPD are linear in $u$ for $u<-32$ dBm, where $-32$ dBm is the largest threshold value for which $R^{2}>0.95$ for all $r>15$. As a result, considering the intersection of the results obtained by Figs.~\ref{fig:meanresvsth}, \ref{fig:modscalevsth}, and \ref{fig:shapevsth}, the optimum values of $u$ and $r$ are $-32$ dBm and $16$, respectively. Additionally, Fig.~\ref{fig:ppqq} shows the probability plots where the black line is the diagonal line to graphically determine the goodness of fit of the GPD. Both PP and QQ plots illustrate that the generalized Pareto model applied to the received power properly follows the empirical results.

We summarize the specifications of Pareto models for other groups in Table~\ref{tabel:parameters}, where $u_{(t)}$, $\xi_{(t)}$, $\tilde{\sigma}_{(u,t)}$, and $l_{GPD}$ denote the optimum threshold in dBm, shape and scale parameters corresponding to $u_{(t)}$, and the log likelihood of GPD, respectively.
\begin{table}[h]
\fontsize{10}{12}\selectfont
    \caption{Estimated time-varying parameters of GPD for different groups.}
\normalsize
    \centering
    \begin{tabular}{c | c| c| c| c} 
    {Group} & $u_{(t)}$ & $\xi_{(t)}$ & $\tilde{\sigma}_{(u,t)}$ & $l_{GPD}$ \\ [0.5ex] \hline\hline
    \textbf{Static car} & $-6$ & $0.126$ & $0.63$ & $-27.55$\\[0.5ex]
    \textbf{Smooth road} & $-24$ & $-0.457$ & $9.21$ & $-55.25$\\[0.5ex]
    \textbf{Ramp road} & $-32$ & $-0.284$ & $8.08$ & $-53.30$ \\[0.5ex]
    \end{tabular}
    \label{tabel:parameters}
\end{table}
\vspace{-0.4cm}

\subsection{Final Channel Model for the Engine Compartment}\label{sec:deviance_engin}
According to Section~\ref{sec:deviance}, the deviance statistic between the GPD models under non-stationary and stationary assumptions $D=2(-136.1-(-1343)) >> 13.28$, where $-136.1$ is the summation of log-likelihoods of \textit{Static car}, \textit{Smooth road}, and \textit{Ramp road} in Table~\ref{tabel:parameters}; $-1343$ is the log-likelihood of the GPD fitted to the whole data sequence, assuming that the channel is stationary \cite{Mehrnia2020}; $13.28$ is the $0.99$ quantile of the $\chi^{2}_{4}$, while $4$ in $\chi^{2}_{4}$ is the difference between the complexities of GPD models under stationary and non-stationary channels which are $2$ and $6$, respectively. To minimize the probability of wrong decision, we assume $\alpha$  as small as possible, i.e., $0.01$, in D-statistic formula. \textcolor{black}{The $D$ statistic result implies that the change-point trend of the threshold, scale and shape parameters of GPD explains a substantial amount of the variation in the channel data sequence while outperforming the GPD model under stationarity assumption.} 

\section{CONCLUSIONS}
\label{sec:conclusions}
In this paper, we introduce a novel framework based on EVT with the goal of modeling the extremes of a non-stationary channel for URLLC. The proposed methodology for modeling a non-stationary channel achieves significantly better fit to the empirical data than the modeling approach for a stationary channel.
\textcolor{black}{In the future, we are planning to extend the proposed framework for the EVT analysis of the non-stationary processes to include a more extensive set of parameters that affect stationarity, such as vehicle speed, road material, and temperature.} \textcolor{black}{Also, the extension of this work for deriving the mathematical expressions of the BER and PER due to the outage based on the proposed model is subject to future work.} 

\balance 

\ifCLASSOPTIONcaptionsoff
  \newpage
\fi

\bibliographystyle{ieeetr}
\bibliography{NonStationaryEVT}

\end{document}